\bigskip
\documentclass[twocolumn,preprintnumbers,amsmath,amssymb]{revtex4}
\usepackage{graphicx}% Include figure files
\usepackage{fancyhdr}
\usepackage{dcolumn}% Align table columns on decimal point
\usepackage{bm}
\usepackage{amssymb}
\usepackage{amsmath}
\usepackage{graphicx}
\usepackage{float}
\usepackage[normalem]{ulem}
\usepackage[dvips]{color}
\usepackage[colorlinks=true,citecolor=blue,linkcolor=black]{hyperref}

\begin{document}
\preprint{ }
\title{ Nuclear potentials relevant to the symmetry energy in chiral models}
\author{  Niu Li$^1$, Si-Na Wei$^2$, Wei-Zhou Jiang$^1$\footnote{Corresponding author, wzjiang@seu.edu.cn}}
\affiliation{$^1$ School of Physics, Southeast University, Nanjing
211189, China\\
$^2$ School of Physics and Optoelectronics, South China University of Technology, Guangzhou 510640, China}
%\date{}
\begin{abstract}
 We employ the extended Nambu-Jona-Lasinio, linear-$\sigma$  models, and the density-dependent  model with chiral limits to work out the mean fields and relevant properties of nuclear matter. To have the constraint from the data, we reexamine the Dirac optical potentials and symmetry potential based on the relativistic impulse approximation (RIA).  Unlike the extended NJL and  the density-dependent models with the chiral limit in terms of the vanishing scalar density, the extended linear-$\sigma$ model with a sluggish changing scalar field loses the chiral limit at the high density end. The various scalar fields can characterize the different Schr\"odinger-equivalent potentials and  kinetic symmetry energy in the whole density region and the symmetry potential in the intermediate density region. The drop of the scalar field due to the chiral restoration results in a clear rise of the kinetic  symmetry energy. The chiral limit in the models gives rise to the softening of the symmetry potential and thereof the symmetry energy at high densities.

\end{abstract}
\maketitle

\section{INTRODUCTION}

Besides the development of the various many-body theories based on the boson exchanges in the quantum field theory, e.g., see Ref.~\cite{BR90}, the gauge invariance is regarded to be important to construct the model for the strong-interacting systems. However, the zero-mass gauge bosons are required by the gauge invariance, which becomes a  puzzle of the Yang-Mills fields when applying to the realistic interacting systems. The success of the Bardeen-Cooper-Schrieffer theory for superconducting electrons~\cite{BCS57} brings the enlightenment that the ground state or  vacuum of the interacting systems is not necessary to respect the gauge symmetry. The symmetry invariant models can possess charges, the temporal component of the currents, that break the vacuum (the vacuum is not annihilated by the  charge).  By early 1960's, it comes to a surge of such model constructions. The typical models are the linear-$\sigma$~\cite{SCh57,GL60} and Nambu-Jona-Lasinio (NJL)~\cite{NJL61} models where the order parameters for the chiral symmetry are the scalar condensate $<\bar{\psi}\psi>$ or the scalar field $\sigma$. In these models, the potentials are characteristic of the field term of the fourth power that ensures a double-well potential, and the chiral parter is the so-called Nambu-Goldstone boson, the pion that is the fundamental and unique boson in the effective field theory~\cite{We79}.  These models provide some clue to the solution of zero-mass puzzle of the Yang-Mills fields. It is interesting to note that Schwinger has the view that the gauge boson does not necessarily have zero mass for the special vacuum~\cite{SCh62}, and soon later  Anderson adds that the gauge boson and Nambu-Golstone boson can cancel each other to leave the finite mass boson only~\cite{AND63}. These are, in fact, the looming prelude for the Higgs mechanism. This, seemly digressive but instead, restates the importance of the vacuum.   Since the original linear-$\sigma$ and NJL models fail to fit  nuclear matter saturation, a dozen of extended models have been developed to fit nuclear matter saturation and properties of finite nuclei~\cite{Bo83,He94,FST97,Sa00,Mi04,Bu05,Zs07,Ts07,Jh08,Hu09,Ja11,We16}. In addition, the chiral symmetry can also be manifest by virtue of the vector channel in the hidden local symmetry~\cite{Br04}. The example of such a vector manifestation of chiral symmetry is the model based on the Brown-Rho (BR) scaling~\cite{Song01,Jia07,Jia07b}. In this work, we single out three models from these categories to expose the role of the chiral symmetry in the nuclear and symmetry potentials relevant to the symmetry energy.

Recently, the uncertainty of the symmetry energy has been again a  hot issue, since the accurate measurement of $^{208}$Pb neutron skin thickness ($0.283\pm0.071$ fm)  through the weak-interaction electron scattering gives a large span of the symmetry energy slope $L=106\pm37$ MeV~\cite{PREX,Re21}. At the same time, a large span of the $42<L<117$ MeV is deduced from the spectra of charged pions~\cite{Es21}. These results seem to shake off the previous constraints on the symmetry energy. In the past, the globally average of 28 independent analyses of various data has led to the value of $L=59\pm 16$ MeV~\cite{Li13}. With inclusion of the lower  ranges either extracted from data~\cite{Lat13,Lat14} or obtained from the ab initio results of neutron matter~\cite{Heb13}, an average of the $L$ values gives  a larger range of $58.7\pm 28.1$ MeV~\cite{Oe16}.
The rising uncertainty of the symmetry energy and the inconsistency in different extractions pose the challenge and meantime the opportunity to study  the symmetry energy from multiple angles. We aim to seek the possible hint and/or constraint on the symmetry energy  by revisiting  the ingredients of the symmetry energy in terms of nuclear potentials in the relativistic  chiral models that in general exhibit the broken vacuum, associated tightly with the symmetry energy through the scalar potential.

We will be interested in  the relativistic impulse approximation (RIA) that  combines the Dirac decomposition of scattering amplitudes
with the nuclear scalar and vector densities.  The direct combination with the scattering data ensures the simplification in the analysis of tangled factors in the strong interaction.  It is known that the optical potentials obtained from the RIA can reproduce the analyzing power and spin-rotation parameter in proton-nuclei scatterings successfully~\cite{m1,m2,l1}, in stark contrast with the standard nonrelativistic optical models~\cite{a11,a22}.   In the past, the RIA  has also been used to study the symmetry
potentials~\cite{Ch05,li06,us2,us3} and in-medium nucleon-nucleon (NN) cross
sections~\cite{Ji07,We21}.  With the help of the RIA,  we study in this work the effects of the chiral symmetry on the  symmetry potential. The chiral models will be solved in the mean-field approximation, and also present the result with  the usual relativistic mean-field (RMF) model for comparison.

The remaining of the paper is organized as follows. In Sec. \ref{RMF}, we briefly
introduce a few typical chiral models and  the RIA. Results and
discussions are  presented in Sec. \ref{results}. A brief summary is
given in Sec. \ref{summary}.

\section{Brief formalism}\label{RMF}
\subsection{Models with chiral symmetry}
The chiral symmetry plays very important roles in   the strong interaction systems.  In the QCD, the chiral
symmetry is cooperated by the gauge symmetry to resolve the axial
anomaly and then point to the more fundamental structure of quarks
and leptons~\cite{Pa75,Hu92}. The chiral symmetry also seems to be a probe to the composite structure of hadrons, which is manifest in the NJL model.  In particular, the chiral phase
transition would generally coincide with the
color deconfinement~\cite{Col80}.

As for  nuclear physics with the broken chiral symmetry, our attention is on the order parameter (the non-vanishing vacuum of the scalar field or condenstate) in chiral models that brings the effects on nuclear potentials, as mentioned in the Introduction. In the following, we interpret simply the extended NJL, linear-$\sigma$ models and the density-dependent  model with chiral limits that are used in this work. The extended NJL model includes the additional interaction terms with the scalar-vector, scalar-isovector couplings ($G_{SV}$, $G_{\rho S} $) to fit the saturation and density dependence of the symmetry energy. Here, the additional terms are given by~\cite{We16,We18,We21}
\begin{eqnarray} \label{NJLQE1}
&&\mathcal{L}_{int}=\frac{G_{SV}}{2}[(\bar{\psi}\psi)^2-(\bar{\psi}\gamma_5\tau\psi)^2] [(\bar{\psi}\gamma_{\mu}\psi)^2+ \nonumber\\
&& (\bar{\psi}\gamma_{\mu}\gamma_5\psi)^2] + \frac{G_{\rho{S}}}{2} [(\bar{\psi}\gamma_{\mu}\tau\psi)^2+(\bar{\psi}\gamma_{\mu}\gamma_{5}\tau\psi)^2]\times\nonumber\\
&&[(\bar{\psi}\psi)^2- (\bar{\psi}\gamma_5\tau\psi)^2].
\end{eqnarray}
The gap equation in the mean-field approximation can be obtained as
\begin{equation}
M^*=m_0-(G_S+G_{SV}\rho_B^2+G_{{\rho}S}\rho_3^2)<\bar{\psi}\psi>,
\end{equation}
where $G_S$ is the scalar coupling, and $m_0$ is the bare nucleon mass.

The linear-$\sigma$ model with additional scalar-vector coupling can have the saturation that leads to a stiff equation of state (EOS) with a very large incompressibility~\cite{Bo83}. Among a large collection of extended models, we choose the one with the following potential obtained from the QCD lattice calculation in the strong coupling limit (SCL)~\cite{Ts07}
\begin{equation}\label{plsm}
V_{SCL}(\sigma)=\frac{1}{2}B_\sigma \sigma^2-A_\sigma\log\sigma^2-C_\sigma\sigma,
\end{equation}
where the coefficients $A_\sigma=f_\pi^2(m_\sigma^2-m_\pi^2)/4$, $B_\sigma=(m_\sigma^2+m_\pi^2)/2$, and $C_\sigma=f_\pi m_\pi^2$. The potential is symmetric about the axis $\sigma=0$ with the minimum of the potential at $\sigma=f_\pi$. The potential in Eq.(\ref{plsm}), instead of the original potential in the fourth power of the $\sigma$, avoids the bifurcation that leads to chiral collapse at the lower chiral condensate. Note that the similar   $\log\sigma^2$ term also appears in a scheme that includes the coupling to the field of the glueball~\cite{He94}.

The density-dependent models with the chiral limit are similar to the simple Walecka model and the density-dependent parameters are based on the BR scaling. The mean-field potential energy is given by~\cite{Jia21}
\begin{equation}
\mathcal{V}=\frac{1}{2} m_\omega^{*2} \omega_0^2 +\frac{1}{2}
m_\rho^{*2} b_{0}^2 +  \frac{1}{2} m_\sigma^{*2}\sigma^2,   \label{eqe1}
 \end{equation}
where  the asterisk on the meson mass denotes the density dependence. The relevant parametrization (SLC and SLCd~~\cite{Jia07,Jia07b}) respects the chiral limit in terms of the vanishing scalar density and nucleon effective mass at high densities. For  usual RMF models, we choose the parametrization Fsugold that contains the nonlinear  self-interactions of the $\sigma$ and $\omega$ mesons~\cite{Fsu}.

\subsection{Relativistic impulse approximation}
\begin{figure}[!htb]
\centering
\includegraphics[height=6cm,width=7cm]{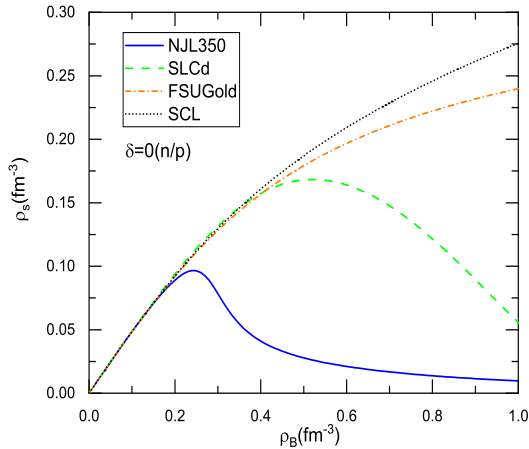}
\caption{The scalar density as a function of
density in  the NJL350, SCL, SLCd, and FSUgold models.}\label{rhos}
\end{figure}

In the proton-nucleus scattering, the scattering process can be
approximately treated   as the incident proton scattered by each of
the nucleons in the target nucleus by neglecting the impact of
the incident particle on the mean fields. The Dirac
optical potential in the RIA can be written as~\cite{m1,m2}:
 \begin{eqnarray}\label{eqria2}
{U}_{\rm opt}=-\frac{4\pi ip_{\rm lab}}{M}
[F_S\rho_{S}+\gamma^0F_V\rho_B] \label{uop},
\end{eqnarray}
where the forward NN elastic scattering amplitudes $F_S$ and $F_V$  are determined directly  from the experimental NN phase shifts~\cite{Ar83}. The RIA optical potential has been used to reproduce pA elastic scattering with incident energies above 400
MeV~\cite{l1} successfully.  $\rho_S$ and $\rho_B$ are the spatial
scalar and vector densities of infinite nuclear matter,
 \begin{eqnarray}
&&\rho_{S,i}=\int_0^{k_{Fi}}\frac{d^3k}{(2\pi)^3}\frac{M^*}{\sqrt{M^{*2}+k^2}},\nonumber\\
&&\rho_{B,i}=\frac{k_{Fi}^3}{3\pi^2},\mbox{ } i=n,p. \label{deo}
\end{eqnarray}
When the density dependent effective mass $M^*$ of nucleons is
obtained from nuclear models, the scalar density can be calculated
from Eq.(\ref{deo}) directly. The scalar density with various models given in the above can give the distinct difference at high densities, as shown in Fig.~\ref{rhos}.  This will give rise to subsequent effects on the optical potentials and relevant quantities.     The Dirac optical potential can be
expressed in terms of  scalar and vector optical potentials:
 \begin{eqnarray}
{U}_{\rm opt}&=&U_S^{\rm tot}+\gamma_0 U_0^{\rm tot},\nonumber\\
U_S^{\rm tot}&=&U_S+iW_S, \mbox{ } U_0^{\rm tot}=U_0+iW_0,
\label{uop2}
\end{eqnarray}
where $U_S$, $W_S$, $U_0$ and $W_0$ are real scalar,
imaginary scalar, real vector and imaginary vector optical
potentials, respectively.

One can derive the Schr\"odinger-equivalent potential (SEP) from the relativistic dispersion relation as ~\cite{Ji07}
 \begin{eqnarray}
U_{\rm sep}^{\rm tot}=U_S^{\rm tot}+ U_0^{\rm tot}+
\frac{{U_S^{\rm tot}}^2- {U_0^{\rm tot}}^2}{2M}+\frac{U_0^{\rm
tot}E_{\rm kin}}{M}, \label{usep1}
\end{eqnarray}
where $U_{\rm sep}^{\rm tot}=U_{\rm sep}+iW_{\rm sep}$ and
$E_{\rm kin}$ is the nucleon kinetic energy.
With the SEP, the symmetry potential is written as
 \begin{equation}
U_{\rm sym}=\frac{U_{\rm sep}^{n}- U_{\rm sep}^{p}}{2\delta},
 \label{usy}
\end{equation}
with $\delta$ being the isospin asymmetry. The $U_{\rm sym}$ is also known as the Lane potential~\cite{mj3}.

\section{Results and discussions}
\label{results}

The models to be used are the extended NJL model (NJL350) with a momentum cutoff 350 MeV~\cite{We16,We18,We21}, SLCd that is a density-dependent relativistic model with chiral limit~\cite{Jia07b,Jia13}, the extended linear-$\sigma$ model in the strong coupling limit (denoted by SCL)~\cite{Ts07}, and the RMF model FSUgold~\cite{Fsu}. The incompressibilities for the models of SCL, NJL350, SLCd, and FSUGold are 279, 262, 230, and 230 MeV, respectively, with corresponding saturation density 0.145, 0.16, 0.16, and 0.145 fm$^{-3}$.

\begin{figure}[!htb]
\centering
\includegraphics[height=6cm,width=7cm]{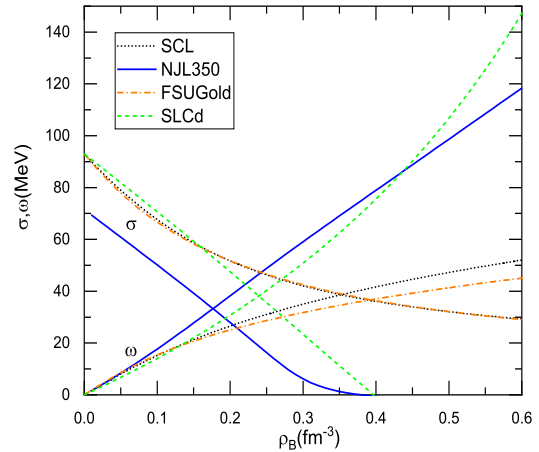}
\caption{The scalar and vector fields. The scalar field is redefined for models SLCd and FSUGold as $f_\pi-\sigma$. }\label{field}
\end{figure}
Figure~\ref{field} shows the scalar and vector fields for these models in the mean-field approximation. As shown in Fig.~\ref{field}, the scalar field with various models can group into two categories: one of which (NJL350 and SLCd) has a fast dropping in the medium, while the other (SCL and FSUGold) has a sluggish decrease. For the chiral models or models with the chiral limit, the scalar field plays a role of the order parameter that reflects the breaking vacuum and probes the chiral restoration in the medium.  For the NJL model, the scalar field is, in fact, equivalent to the scalar condensate $<\bar{\psi}\psi>$. A clear decrease of the scalar field with the NJL350 and SLCd indicates that these two models have the chiral limit in terms of the vanishing of the scalar field or scalar density (see Fig.~\ref{rhos}). The in-medium scalar field with the SCL behaves like that of the RMF model FSUGold, which means that the SCL that respects the chiral symmetry at the vacuum degenerates into  a usual RMF model in the medium. As a result, the SCL is not able to restore the chiral symmetry in dense matter even at the sufficiently high density. This is also  true for other extended linear-$\sigma$ models, as it is not able to bring the scalar field down to vanishing in the medium~\cite{Ts07}. The slowly varying scalar field also provides considerable attraction that softens the  EOS  at high density. Together with rather soft vector potential, the SCL and FSUGold are not able to meet the $2M_\odot$ constraint of neutron stars. The situation with the NJL350 and SLCd is quite different by owning the stiff EOS's at high densities and fitting the $2M_\odot$ constraint of neutron stars~\cite{We16,Jia21,Jia12}.

\begin{figure}[!htb]
\centering
\includegraphics[height=6cm,width=7cm]{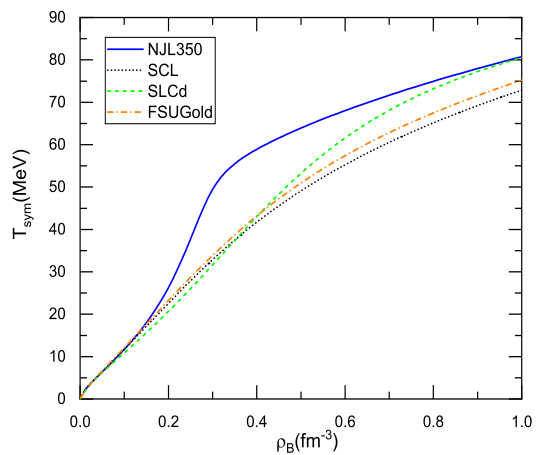}
\caption{The kinetic symmetry energy  as a function of density  with various models.}\label{tsym}
\end{figure}

The scalar field that dictates the nucleon effective mass changes the kinetic symmetry energy (the kinetic part of the symmetry energy) which in the relativistic formulation is $T_{sym}=k_F^2/6E_F$ with $E_F$ being the Fermi energy. The decreasing nucleon effective mass in dense matter increases the $T_{sym}$ clearly. As shown in Fig.~\ref{tsym}, the difference in the kinetic symmetry energy develops beyond saturation density, and it is the most appreciable for the NJL350. With the increase of density, the kinetic symmetry energy with the SLCd comes closer to that with the NJL350 because of  the chiral limit of the SLCd. This is a direct evidence that the (partial) restoration of the chiral symmetry  has a characteristic contribution to the symmetry energy.

\begin{figure}[!htb]
\centering
\includegraphics[height=8cm,width=7cm]{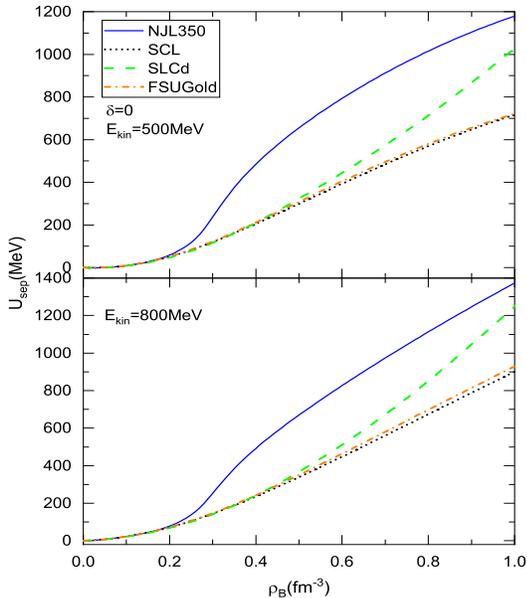}
\caption{The SEP as a function of density at the kinetic energy 500 and 800 MeV.
}\label{usepr}
\end{figure}

Within the framework of the RIA, the nucleon SEP is derived as in Eq.(\ref{usep1}). Figure~\ref{usepr} shows the nucleon SEP of symmetric matter with increasing the density at the two given kinetic energies. The attribution of the difference in various curves is similar to that in Fig.~\ref{tsym} due to the different scalar density. As shown in Fig.~\ref{usepr}, the concrete content of the chiral restoration, such as that reflected by the departure in the NJL350 and SLCd, decides the density dependent behavior of the SEP. With the increase of density, the SEP with the SLCd approaches that with NJL350 due to the increasing eclipse of the nucleon mass. For the models SCL and FSUGold that have a sluggish descent of the nucleon mass, the SEP stays away from those with the models that own the chiral limit. Note that the RIA may not work as well at high densities, and the SEP at the high density end would just be referential.

\begin{figure}[!htb]
\centering
\includegraphics[height=8cm,width=7cm]{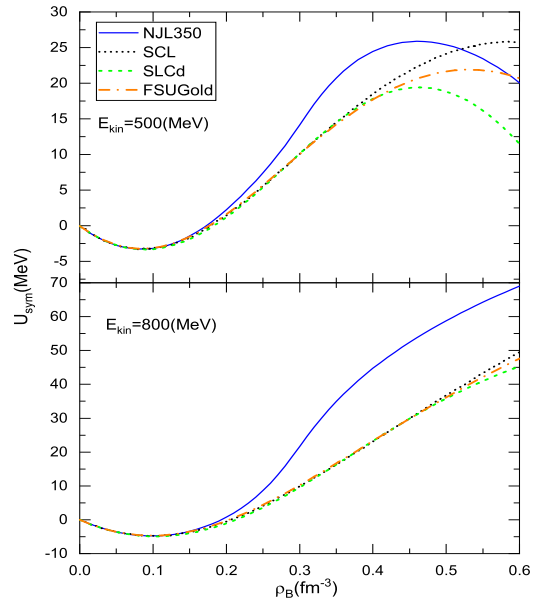}
\caption{The symmetry potential as a function of density at kinetic energy $E_{\rm kin}=500$ and 800 MeV.}\label{usymr}
\end{figure}
With Eq.(\ref{usy}), we can carry out the symmetry potential as  functions of the density or kinetic energy. The symmetry potential is tightly related to the potential part of the symmetry energy~\cite{Ch05,Xu10}. Figure~\ref{usymr} shows the symmetry potential based on various models at $E_{\rm kin}=500$ and 800 MeV. Since at 800 MeV the RIA results actually include some contributions of the inelastic nucleon-nucleon scattering~\cite{We21}, we focus mainly on  the result at $E_{\rm kin}=500$ MeV. We see again that the effect of the  chiral restoration at the intermediate densities (by the NJL350) gives a very clear rise in the symmetry potential. As shown in the upper panel of Fig.~\ref{usymr}, the symmetry potential bends downwards at high densities, which occurs coincidentally in the models NJL350 and SLCd which share the chiral limit. As inferred from Figs.~\ref{tsym} and \ref{usymr}, the symmetry energy appears to be stiffer around saturation density with a larger slope parameter $L$ for the chiral model  which owns the chiral limit. It should be pointed out that the symmetry potential at high densities does not follow in a homeomorphic way  the difference in the scalar density as  those in Figs.~\ref{tsym} and \ref{usepr}. At high densities, there are trespassing between the symmetry potential curves that are subject to the different scalar densities or nucleon effective masses. The reason for this lies in the fact that the variation of the nucleon effective mass against the isospin asymmetry $\delta$ can be alternating in various regions of the nucleon effective mass, yielding the disorder in the SEP and symmetry potential for the nonzero  isospin asymmetry.

\section{Summary}\label{summary}
The chiral symmetry and its breaking  define the vacuum and the Goldstone particles, the pion mesons, for  the non-perturbative strong interaction system and can add restrictions on the nuclear potentials in the medium.
We have revisited the extended NJL and linear-$\sigma$ models which have a nonzero order parameter, the scalar field or the chiral condensate that plays an important role in the properties of bulk matter. Together with the usual RMF model and the density-dependent model with the chiral limit, we have made a comparative study on the nuclear potentials that are relevant to the symmetry energy. It is found  that  the chiral limit in whatever models, chiral or not, ensures a significant reduction of the scalar field and consequently the stiffening of the EOS.  Such a stiffening due to the chiral limit  is also observed in the Schr\"odinger-equivalent potentials and the kinetic symmetry energy. In addition, we find that the models with the chiral limit bring the coincident softening of the symmetry potential at high densities, which suggests the softening of  the symmetry energy at high densities. On the contrary, the sluggish descent of the scalar field makes  the extended  linear-$\sigma$ model to be absent from  the chiral limit in the finite density region and accordingly degenerate into the usual RMF model at high densities.

\section*{ACKNOWLEDGMENT}
The work was supported in part by the National Natural Science
Foundation of China under Grant No. 11775049.

\end{document}